\documentstyle[12pt]{article}   
\begin{document}    
\baselineskip=24.5pt    
\setcounter{page}{1}         
\topskip 0 cm    
\def\lsim{\raise0.3ex\hbox{$<$\kern-0.75em\raise-1.1ex\hbox{$\sim$}}}
\def\gsim{\raise0.3ex\hbox{$>$\kern-0.75em\raise-1.1ex\hbox{$\sim$}}}

\vspace{1 cm}    
\centerline{\Large \bf Resurrection of Grand Unified Theory Baryogenesis}

\vskip 1 cm
\centerline{\large M. Fukugita$^{1}$ and T. Yanagida$^2$}
\vskip5mm
\centerline{$^1$ Institute for Cosmic Ray Research, University of Tokyo,
Kashiwa 2778582, Japan}
\centerline{$^2$ Department of Physics, University of Tokyo, Tokyo 113, Japan}

\centerline{20 March 2002}
 
\vskip 4 cm
\noindent
{\large\bf Abstract}
\medskip

\noindent
A `new' scenario is proposed for baryogenesis. We show that 
delayed decay of coloured Higgs particles in grand unified theories 
may generate excess baryon number of the empirically desired amount,
if the mass of the heaviest neutrino is in the range  
$0.02~{\rm eV}<m_{\nu_3}< 0.8~{\rm eV}$, provided that neutrinos
are of the Majorana type.
The scenario accommodates the case of degenerate neutrino masses, in contrast
to the usual leptogenesis scenario, which does not work when three neutrino
masses are degenerate.

\newpage 
\noindent

After the advent of grand unified theory (GUT) 
the most popular idea for baryon
asymmetry in the universe was to ascribe its origin to baryon number
violating delayed decay of heavy coloured Higgs particles \cite{sakh,gut1,gut2,
gut3}. 
It was later found, however, that standard electroweak theory 
contains baryon
number violation, and this process efficiently erases all baryon numbers
that are produced before the epoch of the electroweak phase transition,
in so far as the baryon excess is produced respecting the $B-L$ conservation
\cite{krs}.
This is the case not only with SU(5) grand unification (and its supersymmetric
extension) but also with any grand unification with higher symmetries that 
has been considered to date. This is because symmetry higher than SU(5) 
contains U(1)$_{B-L}$ as a subgroup, which is unbroken above the grand
unification scale. Even if it is broken at a low energy, 
$\Delta B=\Delta L$ is satisfied
for the excess baryon number in so far as it is generated in decay of
$\phi\rightarrow qq$, $\bar q\ell$ and their conjugate. Multiparticle decays
which would generate baryon numbers with $\Delta B\ne \Delta L$ 
require complicated diagrams and 
are generally too small. In this situation one usually invokes delayed
decay of heavy Majorana particles
with $\Delta L\ne 0$  ($\Delta B=0$) to generate lepton 
number \cite{ leptogen,leptogen2},
and the sphaleron action \cite{krs} to
transfer lepton number to baryon number. 
This mechanism does not particularly require unification of strong 
and electroweak interactions, and it is readily
embedded into many classes of unified theories. 

Experiment has now shown that neutrinos are massive. In particular,
one neutrino that mixes with $\mu$ and $\tau$ neutrinos has a mass
$m_{\nu_3}>0.04$ eV \cite{sk-atm}. In this circumstance we can show that
the GUT scenario using delayed decay of leptoquark Higgs particles is
revived as a possible mechanism of baryon number production, with 
the proviso that neutrinos are of the Majorana type.

We first give a sketch of the idea of a `new' baryogenesis scenario.
The simplest effective interaction that
gives the neutrino the Majorana mass is

\begin{equation}
\frac{1}{2M_{i}}\ell_i\phi \ell_i\phi\ ,
\label{eq:effective}
\end{equation}
where $\phi$ is the standard Higgs doublet and the effective 
mass scale $M_i$ is $<10^{15}$ GeV for $i=3$ from
$m_{\nu_3}>0.04$ eV. With this effective interaction the lepton number
violating interaction $\ell_i+\phi\rightarrow \bar\ell_i+\phi^\dagger$
is in thermal equilibrium above the temperature of $T\sim 10^{14}$ GeV. 
At this high temperature ($T>10^{12}$ GeV) the action of sphaleron 
effects is not effective \cite{sphrate,sphrate2}. Hence, if delayed 
decay of the coloured Higgs particles produces baryon and lepton number 
excess while conserving $B-L$, the lepton number excess is erased by the 
Majorana interaction whereas the baryon number excess is intact. 
When the universe cools to $T<10^{12}$ GeV, the sphaleron action becomes
effective, while the lepton number violating interaction 
already decoupled. This leads to 
baryon number partly converted into lepton
number, conserving $B-L$; 0.35 times the original baryon 
number, however, survives the sphaleron action. The crucial observation
here is that the experimentally indicated neutrino mass points towards
lepton number violation efficiently taking place at very high temperatures
where sphaleron actions are not yet effective, rather than both baryon
and lepton violations undergo at the same temperature, which would
result in vanishing baryon excess \cite{erasure}. 

We now discuss a specific model. 
We consider for simplicity SU(5) GUT, but the model applies straightforwardly
to SO(10) or other GUT without or with supersymmetry. 
We assume the presence of an SU(5) 
singlet ({\bf 1}) fermion in addition to the standard {\bf 5}$^*$ and
{\bf 10} fermions for each family. 
This {\bf 1} may naturally be included in {\bf 16} 
of SO(10). We consider the Lagrangian,
\begin{equation}
{\cal L}_{\rm Yukawa} = 
h^{(k)}_{ij}\psi({\bf 10}_i)\psi({\bf 10}_i)H^{(k)} 
+ f^{(k)}_{ij }\psi({\bf 5^*}_i)\psi({\bf 10}_j)H^{(k)\dagger} 
+ g^{(k)}_{ij}\psi({\bf 1}_i)\psi({\bf 5^*}_j)H^{(k)},
\label{eq:lag}
\end{equation}
where $i=1-3$ and we assume two Higgs particles $k=1,2$ \cite{gut3}.
We suppose that all right-handed Majorana neutrinos, 
$N_i\equiv \psi({\bf 1}_{i}) + \psi^c({\bf 1}_{i})$,
are heavier than the  colour-triplet Higgs particles $H_c$, {\it i.e.} we
have mass hierarchy 

\begin{equation}
10^{12}{\rm GeV}< m_{H_c}<m_N<10^{16}{\rm GeV},
\end{equation}
where the first inequality is the requirement from the limit on
proton instability, and the third inequality is the condition
discussed in what follows. Note that the condition against proton
instability agrees with the energy scale that the sphaleron action
becomes ineffective, i.e., coloured Higgs decay takes place where 
sphaleron effects are not active. 

The effective mass of (\ref{eq:effective}) is given by 
$M_i=m_{N_i}/g_i^2$ with $g_i$ the Yukawa coupling for the right-handed 
neutrino. 
The condition that the reaction rate of 
$\ell_i+\phi\rightarrow \bar\ell_i+\phi^\dagger$,
$\Gamma\approx 0.12T^3/4\pi M_i^2$ be 
sufficiently faster than the expansion
rate of the universe $\gamma_{\rm exp}\approx 17T^2/m_{\rm pl}$
at temperature $T$ is written
\begin{equation}
M_i \lsim 10^{15}\bigg( \frac{T}{10^{14}{\rm GeV}}\bigg)^{1/2}{\rm GeV}.
\label{eq:equil}
\end{equation}
If this inequality is violated for $T\approx m_{N_i}$, the Majorana neutrino
undergoes out-of-equilibrium decay, and lepton excess is generated, 
and the model reduces to the usual leptogenesis scenario.
We require that this does not happen, which
gives $m_{N_i}<10^{16}$ GeV for $g_i\lsim 1$. 
   
We consider the traditional delayed-decay scenario of 
coloured Higgs particles. 
The calculation  
for the baryon abundance in units of specific entropy is
standard \cite{gut3,boltz}. We have

\begin{equation}
{kn_B \over s}\simeq 0.5 \times 10^{-2}\epsilon {1 \over 1+(3K)^{1.2}}\ ,
\end{equation}
where
 
\begin{equation}
K = {1 \over 2}{\Gamma_H\over \gamma_{\rm exp}}\bigg|_{T=m_{H_c}}
  = 3.5\times 10^{17}{\rm GeV}\alpha_H{1 \over m _{H_c}}\ , 
\end{equation}
with  $\Gamma_H\approx\alpha_H m_{H_c}$,
and $\alpha_H=h^2/4\pi$ the Yukawa coupling constant square;
the net baryon number $\epsilon$ produced by pair decay 
of $H_c$ and $\overline{H_c}$ through the interference of one-loop
and tree diagrams is given by

\begin{equation}
\epsilon\approx {\eta\over 8\pi} 10^{-2}\left(F(x)-F(1/x)\right),
\label{eq:eps}
\end{equation}
where $F(x)\simeq 1-x\log[(1+x)/x]$ with $x=m_{H^{(1)}}/m_{H^{(2)}}$ 
the ratio of the masses of
two  coloured Higgs particles and $\eta=\sin\left({\rm arg}[ 
{\rm tr} (f^{(1)\dagger} f^{(2)}h^{(1)\dagger} h^{(2)})]\right)$ 
is the factor representing the CP-violation phase.
To obtain the numerical factor of (\ref{eq:eps}) we use
$h^{(i)}\approx 1$ and $f^{(i)}\approx 0.1$ from masses of quarks.
For example, if we take 
$m_{H_c}\approx 10^{15}$ GeV, $x\approx 0.5$ and 
$\eta\approx 0.1$, we obtain $kn_B/s\approx 2.5\times 10^{-10}$ nominally in
agreement with the empirical baryon abundance. This process produces 
lepton number at the same time by the amount $\Delta L=\Delta B$.

Produced lepton number, however, is erased if the Majorana 
interaction  is in the thermal equilibrium at $T\approx m_{H_c}$.
If we take $m_{H_c}\lsim 10^{15}$ GeV the condition for thermal
equilibrium is read from eq. (\ref{eq:equil}), which leads to 

\begin{equation}
m_{\nu_i}\gsim 2\times 10^{-2}{\rm eV}\ ,
\label{eq:numass1}
\end{equation}
using (\ref{eq:effective}) with $\langle\phi\rangle\simeq 250$ GeV. This
condition should be satisfied at least for one species of neutrinos.

The rate for the action of sphalerons is computed to be \cite{sphrate,
sphrate2}

\begin{equation}
\Gamma_{\rm sph}\approx 2\times 10^2 \alpha_W^5T\ ,
\end{equation}
where $ \alpha_W\approx 1/40$ is the weak coupling constant. 
$\Gamma _{\rm sph}>\gamma_{\rm exp}$ gives $T\lsim 12 \alpha^5_Wm_{\rm pl}
\approx 1.4\times 10^{12}$ GeV for the temperature, below which
the sphaleron action
becomes effective. We must require that the Majorana interaction
decouples by this temperature, or otherwise all existing baryon and lepton 
numbers are erased by the joint action of sphalerons and Majorana
interactions \cite{erasure}. 
The condition obtained from eq. (\ref{eq:equil}) with the aid of
(\ref{eq:effective}) and the value of $\langle\phi\rangle$ is 

\begin{equation}
m_{\nu_j}< 0.8{\rm eV}.
\label{eq:numass2}
\end{equation}
This must be satisfied for all neutrinos.
The action of sphalerons at lower temperatures then partially
converts baryon number to lepton number,
but baryon number remains by the amount of \cite{leftover}
\begin{equation}
\Delta B_f= {{8N_f + 4N_H} \over {22N_f + 13N_H}} \Delta  B_i 
=0.35\Delta B_i\ 
\end{equation}
for three generations of fermion families $N_f=3$ and two Higgs doublets
$N_H=2$. Hence we expect $kn_B/s\approx 1\times 10^{-10}$ 
with the parameters exemplified above.

Our central result is summarized as follows.
If the two inequalities (\ref{eq:numass1}) and (\ref{eq:numass2})
are satisfied, {\it i.e.} if the mass of the heaviest neutrino
satisfies $0.02<m_{\nu_3}<0.8$ eV, baryogenesis via coloured Higgs decay
works within the framework of GUT. 
This neutrino mass range nearly coincides
with the limits derived empirically:
atmospheric neutrino oscillation gives a lower
limit on the $\tau$ neutrino mass, $m_{\nu_3}>0.04$ eV \cite{sk-atm},
and the limit from neutrinoless double beta decay experiment is about
$\langle m_{\nu_1}\rangle <0.5-1.5$ eV \cite{dbeta1}, or $\sum_i m_{\nu_i}<4$
eV from cosmology \cite{cosmol}.

We emphasize that the present scenario is valid with neutrinos 
nearly degenerate in mass. 
This contrasts to the usual leptogenesis scenario of delayed heavy Majorana
neutrino decay, for which 

\begin{equation}
\mu_1=\left( h_{11}{1 \over M_1}h_{11}^\dagger+
h_{21}{1 \over M_2}h_{12}^\dagger+h_{31}{1 \over M_3}h_{13}^\dagger\right)
\end{equation}
must satisfy $\mu_1\lsim 2\times 10^{-3}$ eV. While this mass term
is not directly related with the physical neutrino mass

\begin{equation}
m_{\nu_i}=\left( h_{i1}{1 \over M_1}h_{1i}+
h_{i2}{1 \over M_2}h_{2i}+h_{i3}{1 \over M_3}h_{3i}\right)\ ,
\end{equation}
it is clear that one or two neutrinos must have small masses. Namely,
the neutrino mass must be hierarchical. An additional reason that
disfavours the degenerate neutrino masses for the leptogenesis scenario is
the necessity of hierarchy in the right-handed neutrino masses. If they
are degenerate, produced baryon number vanishes.

The neutrino mass would give a diagnostics as to which baryogenesis
scenario is to be realized.
If a future neutrino mass experiment would prove that the mass of
three neutrinos has some non-zero baseline value in excess of 0.01 eV,
say by a positive detection of neutrinoless double beta decay, 
the Higgs decay scenario given in this paper would be a more promising
possibility for baryon number generation. If the hierarchical neutrino
mass is favoured for some reasons, either of the two baryogenesis scenarios
is equally viable.

\end{document}